\documentclass[aps,prl,floatfix]{revtex4}
\usepackage{epsfig}
\begin{document}


\title{Spin-polarized bipolar transport and its applications}

\author{S. Das Sarma,$^{1}$ Jaroslav Fabian,$^{2}$ and Igor \v{Z}uti\'{c}$^1$} 
\affiliation{$^1$Condensed Matter Theory Center, Department of Physics, University of Maryland, 
College Park, Maryland 20742-4111, USA\\
$^2$Institute for Theoretical Physics, Karl-Franzens University, Universit\"{a}tsplatz 5, 
8010 Graz, Austria}

\begin{abstract}
In spin-polarized bipolar transport both electrons and holes in doped semiconductors
contribute to spin-charge coupling. The current conversion between
the minority (as referred to carriers and not spin) and majority carriers leads to novel spintronic
schemes if nonequilibrium spin is present. Most striking phenomena
occur in inhomogeneously doped magnetic {\it p-n} junctions, where the presence
of nonequilibrium spin at the depletion layer leads to the spin-voltaic effect:
electric current flows without external bias, powered only by spin. The spin-voltaic effect
manifests itself in giant magnetoresistance of magnetic {\it p-n} junctions, where the relative 
change of the magnitude
of electric current upon reversing magnetic field can be more than 1000\%.
The paper reviews nonmagnetic and magnetic spin-polarized {\it p-n} junctions, formulates 
the essentials of spin-polarized  bipolar transport as carrier recombination and spin
relaxation limited drift and diffusion, and discusses specific device schemes
of spin-polarized solar cells and magnetic diodes.
\end{abstract}

\maketitle

\section{Introduction}

Semiconductor spintronics~\cite{dassarma01,dassarma00a,dassarma00b} is a growing field 
where the fundamentals of spin-polarized transport and spin relaxation 
processes~\cite{optical84,fabian99} 
in homogeneous and inhomogeneous semiconductors are being developed, 
and novel electronic device schemes based on active control of spin in semiconductor
heterostructures are
being proposed. Semiconductor spintronic devices now
include various spin field-effect transistors, spin valves and 
spin detectors, and spin bipolar diodes and transistors~\cite{dassarma01}. 
The potential
of semiconductor spintronics is underscored by the 
discoveries
of novel magnetic semiconductors (some of which are reported to 
remain ferromagnetic even at room 
temperature~\cite{CdMnGeP,reed02,hebard02,matsumoto01,cho02}), 
which can be successfully incorporated
with nonmagnetic semiconductors to extend the functionality of traditional
semiconductor devices. Furthermore, it has been demonstrated that the
ferromagnetism in semiconductors can be controlled both 
optically~\cite{koshihara97,oiwa02} and electrically~\cite{ohno00,park02}.

While some of the interest, stimulated by possible applications, to actively manipulate 
spin in semiconductors has only emerged in recent years, theoretical understanding for spin 
manipulation in electronic materials has been steadily accumulated over the last several 
decades~\cite{dassarma00a,optical84,fabian99,rashba02a,zutic02}.
Often the concepts currently applied to spintronics schemes involving
semiconductors have been initially investigated in other materials systems
or in simpler nonmagnetic homogeneous semiconductors.
Crucial for spintronics is spin-charge coupling, where spin degrees
of freedom influence transport of charges and, {\it vice versa}, where
spin transport depends on nonequilibrium charges. Spin-charge coupling
appears at a microscopic level, where individual electron dynamics
depends on the electron spin orientation, the fact usually expressed by
the spin-orbit  
term in the Hamiltonian. Another form of spin-charge
coupling occurs in statistical ensembles of electrons where spin and
charge are coupled because electrons carry both spin and charge and because 
the Pauli principle limits the occupation of the states to single spins.
As a result, the transport properties of an electronic system are greatly influenced
by spin polarization, since electrons taking part in transport 
will have different densities of states, Fermi velocities, scattering
times, etc., for different spin orientations. The microscopic nature of 
spin-charge coupling is exemplified by the Rashba effect~\cite{rashba60,rashba84}
occurring typically 
in GaAs-like  semiconductor 
interfaces, where the spin-orbit interaction and thus the spin-charge
coupling in the system 
can be effectively controlled by electric fields (say from a 
biasing gate~\cite{nitta97,miller02}).
Additional studies have examined the Rashba spin-orbit interaction in many other systems (
for example, near the interfaces with superconductors~\cite{gorkov01} and in interacting
quantum wires~\cite{hausler01}). 
The statistical spin-charge coupling 
was initially suggested by Aronov and Pikus~\cite{aronov76a,aronov76b}.
They were the first to realize that electrical spin injection 
is possible: the 
charge current flowing from a ferromagnetic electrode can give rise to nonequilibrium
spin in the connected paramagnetic metal or semiconductor. The reverse process, that 
of generating charge current (or emf in an open circuit) by nonequilibrium spin, was first described 
by Silsbee~\cite{silsbee80}, and later demonstrated experimentally and developed further theoretically
by Johnson and Silsbee~\cite{johnson85,johnson87}, to serve the purpose of detecting the 
amount of the injected spin by electrical means.
The Silsbee-Johnson spin-charge coupling was realized  as
the phenomenon where the presence of nonequilibrium (ensemble) 
spin influences charge properties of the interfacial system. In particular, the Silsbee-Johnson
spin-charge coupling states that an emf appears across the interface of
a magnetic and a paramagnetic metal, if nonequilibrium magnetization
is maintained (by external spin injection, for example) in the paramagnetic metal.

Rashba's ideas significantly influenced spintronics by giving rise to the
concept of the spin field-effect transistor of  Datta and Das~\cite{datta90}. 
This proposal stimulated an extensive body of
work on spintronic devices and ballistic and mesoscopic spin-polarized
transport. Our own work on spin-polarized  bipolar transport--what we call
{\it bipolar spintronics}--drew from the ideas of Silsbee and Johnson on spin-charge
coupling as a statistical notion. The motivation we had in mind was to examine the effect 
of both electrons and holes on spin-polarized transport in inhomogeneous semiconductors
and to propose simple device 
structures~\cite{dassarma00c,zutic01a,zutic01b,zutic02a,zutic02b,fabian02a,fabian02b} where
possible applications of bipolar spintronics can be studied.
One of  
the devices we consider, the magnetic
{\it p-n} junction~\cite{zutic02a,fabian02a}, is a generalization of the Silsbee-Jonhson 
coupling to
nondegenerate semiconductor interfaces (essentially depletion regions). 
When the minority region of a {\it p-n} junction is magnetic (that is, it 
has a significant, on the level of thermal energy, equilibrium spin band splitting
of the minority carriers) 
and the corresponding majority region is nonmagnetic, electric current will flow through the 
junction if nonequilibrium spin is maintained at the depletion layer in the 
nonmagnetic majority side. This is the spin-voltaic effect~\cite{zutic02a,zutic02b,fabian02a} 
in magnetic {\it p-n} junctions, which is also manifest
as a giant magnetoresistance of the
junction when forward bias is applied.

In this paper we first review the Silsbee-Johnson concept of spin-charge
coupling, then introduce the equations for spin-polarized  bipolar
transport in inhomogeneous magnetic semiconductors, and finally discuss
two specific bipolar spintronic schemes: spin-polarized {\it p-n}
junctions~\cite{zutic01a,zutic01b} and magnetic diodes~\cite{zutic02a,zutic02b,fabian02a}.

\section{Silsbee-Johnson spin-charge coupling}

As recently emphasized by Rashba, spin emf is a specific manifestation of the general principle
that macroscopic (dissipative) currents flow in nonequilibrium 
situations~\cite{rashba02a,rashba02b,landau36}.
Since both spin and charge are carried by electrons,
nonequilibrium spin can give rise not only to spin current, but to electric
current (or emf, properly called spin emf, in an open circuit) as well. 
This is a purely statistical property of the charge and spin
carriers, not of the microscopic dynamics of electrons according to a spin-dependent
Hamiltonian (of the Rashba type, for example). Spin-dependent dynamics manifests
itself here only on the macroscopic level in limiting the amount of the accumulated 
nonequlibrium spin and in the conversion of the spin energy into thermal energy in the
process of spin relaxation. 

In the Silsbee-Johnson effect the ferromagnetic
and paramagnetic degenerate electron systems (metals or degenerate semiconductors)
form an electrical and spin contact. Generating nonequilibrium spin in the paramagnetic system
makes, say, spin up electrons flowing into the ferromagnet, and, vice versa,
spin down electrons (if the spin down subband is not fully occupied) flow in 
the opposite direction. These two currents are not in balance, due to the unequal
conductances of the two spin species in the ferromagnet, resulting in a charge current. 
If the two systems were paramagnetic metals, no charge
current would flow, since equal currents would flow in both directions. The only 
flowing current (this is a nonequilibrium situation after all) would be spin 
current. The Silsbee-Johnson effect is also a spin-valve effect, since the electric current direction 
can be reversed by reversing either the nonequilbrium spin in the paramagnet or
the equilibrium magnetization of the ferromagnet. 

Silsbee and Johnson succeeded to measure the induced open circuit spin emf of the order
of pico volts, which is equivalent to measuring one spin per 10$^{11}$~\cite{johnson85}.
To obtain the value of spin emf, consider for simplicity the ferromagnet to be an
ideal Stoner ferromagnet with only the minority (say, spin up) spin electrons 
participating in current conduction
(the majority spin subband is fully occupied). Maintaining nonequilbrium spin in the
paramagnet means maintaining a difference in the chemical potentials between the two 
spins. This difference leads to charge current across the interface, proportional to the
difference between the chemical potential of spin up electrons in the paramagnet
and in the ferromagnet. In an open circuit, the current does not flow and the chemical potential
for spin up electrons will be constant across the interface, which will be achieved
by the build up of the spin emf $V_{\rm spin}$. The spin emf is simply the difference between the
chemical potentials when current flows. This difference can be written for degenerate
statistics as $\delta \mu\equiv eV_{\rm spin}=\delta M/\mu_B g_F$, where $\mu_B$ is the
Bohr magneton, $\delta M$ is the nonequilbrium magnetization in the paramagnet, and
$g_F$ is the density of states in the paramagnet at the Fermi level. Considering that
for degenerate electron gas paramagnetic susceptibility is $\chi=\mu_B^2 g_F$, the
spin emf becomes
\begin{equation}\label{eq:spinemf}
V_{\rm spin}=\frac{\mu_B \delta M}{e \chi}.
\end{equation}
For a realistic ferromagnet in which both spin states are conducting the right 
hand side of Eq.~\ref{eq:spinemf} is multiplied by a suitably defined spin 
polarization \cite{johnson87}.
 
The described spin-charge coupling can be used to detect nonequilibrium spins at
magnetic-paramagnetic interfaces, and, when combined with spin injection, inquire
about spin diffusion lengths in the paramagnet without the need to apply magnetic
field. We will show below that a nonlinear (in bias voltage) analogue of the spin-charge
coupling in degenerate systems exists in nondegenerate semiconductors, where the role
of the interface is played by the depletion layer sandwiched between a magnetic and
nonmagnetic semiconductor forming a {\it p-n} junction.

\section{Spin-polarized bipolar transport}  

Spin-polarized  bipolar 
transport refers to transport in doped semiconductors
where both electrons and holes contribute to charge and spin currents. Usefulness
of the concept of spin-polarized bipolar transport is 
demonstrated in 
microelectronic device  
schemes, most prominently in novel magnetic {\it p-n} junctions which
are formed by inhomogeneously doped--both with donors and acceptors, and with magnetic
impurities inducing large carrier $g$-factors--semiconductors. Traditional {\it p-n}
junctions~\cite{tiwari92} are the basis for all electronic (and part of optical) information technology,
since they form both bipolar junction transistors and unipolar MOSFETs, as well as
photodiodes, light emitting diodes, and diode lasers. It is a great challenge for spintronics to 
take advantage of the existing semiconductor technology and extend its capabilities, 
for example,
to build on-single-chip computer logic processors and nonvolatile random access
memory. We believe that magnetic diodes may prove fundamental devices also in 
semiconductor spintronics.

Nonequilibrium spin can be injected into semiconductors and then be transported by
drift and diffusion throughout the sample. The transport is limited by spin relaxation.
If, in addition, both electrons and holes form the current, the transport
is further limited by carrier (electron and hole) recombination. This bipolar transport
is particularly important in {\it p-n} junctions, where charge current depends on 
how fast electrons and holes recombine (in other words, on how fast the minority, say
electron current, is converted to the majority hole current in the $p$ region
of the junction). The presence of nonequilibrium
spin either enhances the current by helping building up the minority carrier
density, or inhibits the current by reducing the minority carrier density. We will
demonstrate this response of charges to nonequilbrium spin in magnetic {\it p-n} junctions 
in a subsequent chapter. 

The general drift-diffusion  
equations for bipolar carrier and spin  transport in inhomogeneous
magnetic semiconductors were first written in Ref.~\cite{zutic02a}. Several
limiting cases of these equations, for a homogeneous magnetic~\cite{martin02}
and nonmagnetic~\cite{yu02} semiconductor, have been subsequently investigated.
For illustration,  consider only electrons to be 
spin polarized, leaving holes spin unpolarized. Denote as $n$, $p$, and $s$
electron, hole, and (electron) spin densities. Electron ($J_n$) and spin ($J_s$) currents
are~\cite{fabian02a} 
\begin{eqnarray}
J_n&=&D(n\phi' + s\zeta'-n'), \\
J_s&=&D(s\phi' + n\zeta'-s'),
\end{eqnarray}
where $D$ is electron diffusivity, $\phi$ is the electric potential (from the
applied bias and built-up internal charges), and $\zeta$ is the equilibrium
spin splitting of the conduction band (arising from the large electron $g$ factor
and applied magnetic field). The  first terms on the right hand sides are the electric 
drift, the second terms are the magnetic drift, while the last terms represent electron
and spin diffusion arising from nonuniform electron and spin densities. In this
paper we write the potentials in the units of $k_B T/q$, where $k_B$ is the
Boltzmann constant, $T$ is temperature, and $q$ is proton charge. The drift
and diffusion of electrons and spin are limited by electron-hole recombination
processes, as well as by spin relaxation processes. This limitation can be
expressed by continuity relations, which in a steady state read
\begin{eqnarray} 
J_n'&=&-r(np-n_0p_0), \\
J_s'&=&-r(sp-s_0p_0)-\frac{s-\alpha_0n}{T_1},
\end{eqnarray}
where $r$ is the constant of the electron-hole recombination rate and $T_1$ is the
spin relaxation time. The subscript zero denotes the equilibrium values (when
no external bias is applied and spin is in equilibrium). Spin polarization is
denoted here as $\alpha\equiv s/n$.  Together with Poisson's equation relating
the built-up charge density to electric potential $\phi$, the above equations for carrier
and spin drift and diffusion, and carrier recombination and spin relaxation fully
describe the steady state density and current profiles of carriers and spins
in inhomogeneous magnetic semiconductors. The equations cannot be solved exactly
analytically, and one must resort to numerical methods for general cases. We have solved the
equations for a few cases of interest, including spin-polarized {\it p-n} 
junctions~\cite{zutic01a}, spin-polarized solar cells~\cite{zutic01b}, 
and magnetic {\it p-n} junctions~\cite{zutic02a}. We have also succeeded
in solving the equations analytically for the general case of spin-polarized magnetic
{\it p-n} junctions at low biases~\cite{fabian02a}, 
the regime in which most devices based on {\it p-n} junctions operate. 
In the subsequent sections we describe 
several elementary bipolar spintronic devices,
and illustrate their properties using mainly the results of our numerical calculations.

\section{Bipolar spintronic devices}

The most fundamental bipolar spintronic device is a spin-polarized magnetic {\it p-n}
junction. Such a device in  equilibrium has, in general inhomogeneous,
spin splitting of the carrier bands, and the nonequilibrium spin can be
externally injected. The equilibrium spin splitting can be caused,
for example, by doping the junction inhomogeneously with magnetic impurities and placing
the junction in a magnetic field. Another possibility is to use ferromagnetic semiconductors
(ferromagnetic at a specific donor or acceptor doping density to achieve inhomogeneity).
Since spin-polarized magnetic {\it p-n} junctions form quite a general group of devices,
it is useful to consider particular cases of interest. Below we consider spin-polarized
nonmagnetic {\it p-n} junctions,
which are just ordinary {\it p-n} junctions but with 
nonequilibrium spin added (by illumination of circularly polarized light or external 
spin injection, for example) and spin-polarized magnetic {\it p-n} junctions in which
one of the regions, $p$ or $n$, is magnetic in equilibrium, but the equilibrium spin
splitting is constant throughout this specific region. Fascinating phenomena occur also
in {\it p-n} junctions in which the magnetic region itself is magnetically inhomogeneous
(as described in Ref.~\cite{fabian02a}), and where magnetic drift plays active role in determining 
charge transport. We will not discuss such devices here, due to their added complexity.
What all the spin-polarized (magnetic and nonmagnetic {\it p-n} junctions) have 
in common is the interplay between charges and spins, and the fundamental concept which
we have formulated, namely the spin injection through the depletion layer. Bipolar spintronics
is useful only so far as spins can be injected from the majority region into the minority
one (and {\it vice versa}) through the depletion layer. We will illustrate below how this
concept of spin injection is realized in the two cases of nonmagnetic and magnetic 
spin-polarized {\it p-n} junctions. 

\subsection{Spin-polarized p-n junction}

\begin{figure}
\centerline{\psfig{file=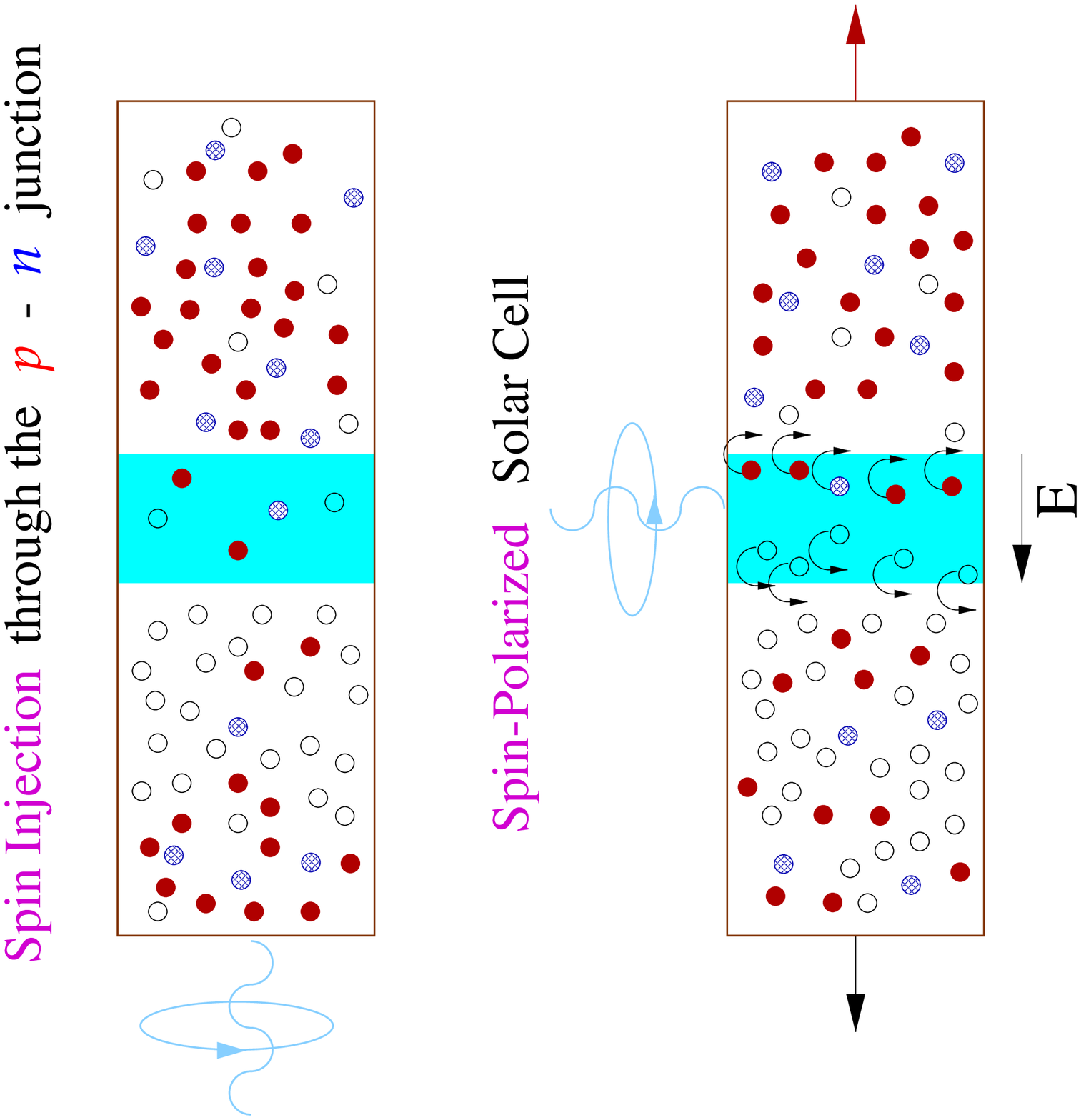,width=1\linewidth,angle=-90}}
\caption{ Top: Spin injection through a {\it p-n} junction can be observed
by shining circularly polarized light at the $p$ region, orienting
spins of the minority electrons, which then carry the spin into the
majority $n$ region. The result is spin pumping of the majority
carriers by the minority ones, an analogue of optical spin pumping
of majority carriers in semiconductors. 
Holes are denoted as empty circles while majority and minority spin
electrons are represented by filled and patterned circles, respectively.
Bottom: Spin-polarized solar
cells use circularly polarized light shining on (or close to) the depletion
layer. Spin-polarized electrons from the depletion layer are swept to the
majority $n$ side by the huge built-in field of the junction, resulting 
in both spin and charge currents through the junction.  
 }
\label{fig:1}
\end{figure}

A spin-polarized {\it p-n} junction is a {\it p-n} junction where spin-polarized
current flows as a result of external spin excitation. We have considered the cases
where nonequilibrium spin is created by shining circularly polarized light on one
of the ends of the junction (see Fig.~\ref{fig:1}). In all our simulations we keep only electrons spin 
polarized and the externally injected (what we call source) spin was implemented as a boundary condition
in the drift-diffusion equations. The fundamental question of whether source
spin can be further transported across the depletion layer has positive answer. Both
spin injection from the $p$ region to the $n$ region, and vice versa, from the $n$ region
to the $p$ region are very effective. 

\begin{figure}
\centerline{\psfig{file=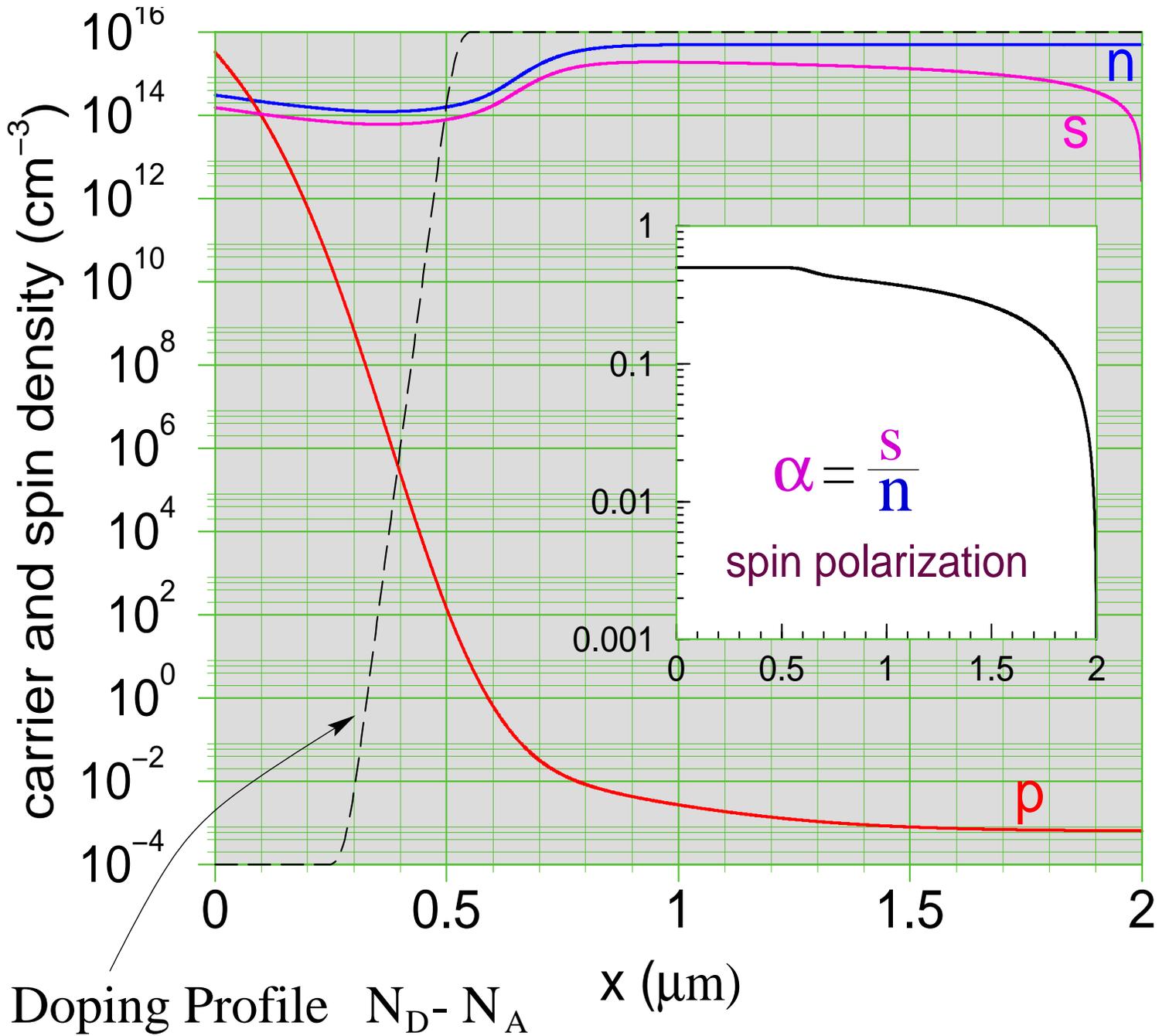,width=1\linewidth,angle=-90}}
\caption{Numerical results~\cite{zutic01a} for spin injection through the {\it p-n} junction. Electron
and hole densities $n$ and $p$ are
shown as a function of the distance $x$
from the illuminated surface ($x=0$).
The dashed line is the doping profile $N_D-N_A$ (scale not shown): it is
$p$-type with $N_A=3\times 10^{15}$ cm$^{-3}$ on the left and $n$-type with
$N_D=5\times 10^{15}$ cm$^{-3}$ on the right; the transition region is between
$0.25$ and $0.55$ $\mu$m.
Also plotted is the spin density $s=n_\uparrow-n_\downarrow$ and the spin
polarization $\alpha=n/s$ in the inset. The remarkable result that $\alpha$
extends well beyond the transition layer (and $s$ is amplified) demonstrates
both spin injection and spin
density amplification.
}
\label{fig:2}
\end{figure}

If source spin appears in the $p$ region, this means that the minority electrons are
spin polarized. Since both carrier (here synonymous for electron) 
and spin densities are out of equilibrium 
if light illuminates the sample in the $p$ region, the electrons carrying
spin diffuse towards the depletion layer where they are swept by the large
built-in field to the $n$ side, where they form the majority carriers. In the
$n$ region the spin can further diffuse away from the depletion layer, or
relax by $T_1$ processes. If the light illumination creates $G$ electrons per
second per unit volume, with spin polarization $\alpha$, the spins at the
depletion layer in the $p$ side appear with the rate of roughly $\alpha G$. 
Considering that all the electrons (and thus spins) arriving at the depletion
layer are swept to the $n$ side, this rate must equal, in a steady state, the
rate of spin relaxation and spin diffusion in the $n$ side. A typical scale for these
rates is $s/T_1$. As a result, the nonequilibrium injected spin in the majority
region is $s \approx \alpha G T_1$ (more complete considerations better theory 
give 
$s=\alpha G \sqrt{T_1 \tau_s}$, 
where $\tau_s$ is the effective spin relaxation time in the minority $p$ region).
 This is normally larger than the spin 
in the minority region itself, an effect we called spin amplification~\cite{zutic01a}. 
The descriptive name for the effect is spin pumping of the majority carriers
by the minority channel, since it is analogous to spin pumping by circularly 
polarized light of majority carriers in semiconductors. Our numerical 
calculation demonstrating the effect is given in Fig.~\ref{fig:2}. 


One application
of a spin-polarized {\it p-n} junction is a spin-polarized  solar cell (Fig.~\ref{fig:1}).
If circularly polarized light illuminates the junction close to the depletion
layer, nonequilibrium charges and spins can arrive in the layer and be powered
by the equilibrium built-in electric field to form a spin-polarized current. This
could be useful in generating spin-polarized currents~\cite{zutic01b}. 
Another use of spin pumping
by the minority carriers is in significantly extending the effective spin diffusion 
length (again the effect of the built-in field) and in making it possible to 
control, by an applied reverse bias to the junction, the amount of the injected
nonequilibrium spin~\cite{zutic01a}. The width of a 
depletion layer changes with the applied bias, which changes
the amount of spin arriving in the $p$ side at the depletion layer. This significantly
affects the amount of injected spin. We called this effect the spin capacitance 
effect~\cite{zutic01a}, since it is an analogue of the voltage to charge relation in capacitors.

The question of injected spin across the depletion layer from the majority
(here $n$) side to the minority one is of equal importance, since diodes
operating under forward biases would be able to carry the spin information 
from one lead to the other across the junction. Injecting spin from the majority to the minority
side is also important for potential applications of spintronics in bipolar
junction transistors~\cite{zutic01a,fabian02b}, where two {\it p-n} junctions would operate back-to-back
and both spin injection regimes would be needed for injecting spin all the way through
the transistor, as well as  for amplifying the spin-polarized current from the emitter
to the collector. Our numerical calculations\cite{zutic01a,zutic01b} demonstrate
that spin injection by the majority carriers is very effective. Spin polarization
does not change significantly upon entering the $p$ region. As we have shown
analytically~\cite{fabian02a}, the {\it nonequilibrium} spin polarization is actually uniform across the
depletion layer. So whatever spin-polarization arrives at the depletion layer from the
$n$ region, it must appear across in the $p$ side.

\subsection{Magnetic diode}

A magnetic {\it p-n} junction would typically contain magnetic impurities
to amplify its magnetic response. An applied magnetic field would lead to 
significant (on the scale of the thermal energy) spin 
splitting~\cite{furdyna88,dietl94,fiederling99,jonker00} of carrier bands in  
such a junction. Here we consider a particular case where only one of the
regions is magnetic, and the equilbrium spin splitting is constant within
the region (that is, the spin splitting varies only in the transition region
of the junction). The fundamental question of spin injection through the
depletion layer is more complicated than in nonmagnetic junctions. It would
be natural to expect that if, say, the $n$ region is magnetic and the junction
is under forward bias, spin injection would occur since spin-polarized electrons
from the $n$ region carry the spin into the minority side, similarly to
what happens in ferromagnet-paramagnetic metal spin injection. We
have demonstrated rather that spin injection in magnetic {\it p-n}
junctions can occur only at large biases, beyond the usual region
of interest for devices. At small biases, where the junctions show their
typical exponential rectification I-V characteristics, spin injection is
not possible. The reason is as follows. Let there be more spin up than spin
down electrons in the $n$ region. The current electrons with a particular 
spin orientation across
the depletion layer depends on the number of electrons available for
transport and on the barrier height for thermal transport across the
junction's depletion layer. This barrier is higher for spin up than
for spin down electrons, exactly compensating for the greater density of
the former. As a result, spin current will not flow across the depletion
layer and spin will not be injected. This is the consequence of the 
fact that at low biases the depletion layer can be considered to be
(quasi)equilibrium, and only nonequilibrium spin can be injected through
the depletion layer. Once
the bias is increased, nonequilibrium spin accumulates already in the
magnetic region, and the spin is then injected thought the depletion
layer to the minority nonmagnetic side.  The corresponding numerical
calculation is shown in Fig.~\ref{fig:4}, where we also discuss the
opposite case of spin injection from the minority magnetic side. There
spin injection appears rather as  spin extraction, and contains
the same physics as large bias spin injection from the magnetic majority
side.


Magnetic {\it p-n} junctions are potentially useful for their nonlinear spin-charge
coupling. This coupling appears at low biases if nonequilibrium source spin
is introduced into the junction. Spin polarization can then be injected
through the depletion layer without difficulties. In addition, the nonequilibrium
spin significantly affects charge transport across the junctions. Consider
the case when the minority $p$ region is magnetic and the majority $n$ region
is nonmagnetic, but a source spin is externally injected into the region (either by 
illuminating with a circularly polarized light or by a third-terminal spin
injection~\cite{zutic02b,fabian02a}). As the source spin diffuses towards the depletion layer, it 
will come into contact with the equilibrium magnetization of the minority
region. The nonequilibrium spin perturbs  
the balance between the generation
and recombination currents across the depletion layer, leading to a net charge
current. Remarkably, this current flows even if the external biasing battery
is switched off. The current is powered by nonequilbrium spin, in analogy
with the current across the ferromagnet-paramagnetic metal in the Silsbee-Johnson
effect. And similarly to the Silsbee-Johnson effect, the charge current can 
change its direction by reversing the external magnetic field (that is, by
reversing the equilibrium polarization of the magnetic side) or by reversing
the orientation of the source spin. This is the spin-voltaic and the spin valve
effect of magnetic {\it p-n} junctions introduced in 
Refs.~\cite{zutic02a,zutic02b,fabian02a}. 

\begin{figure}
\centerline{\psfig{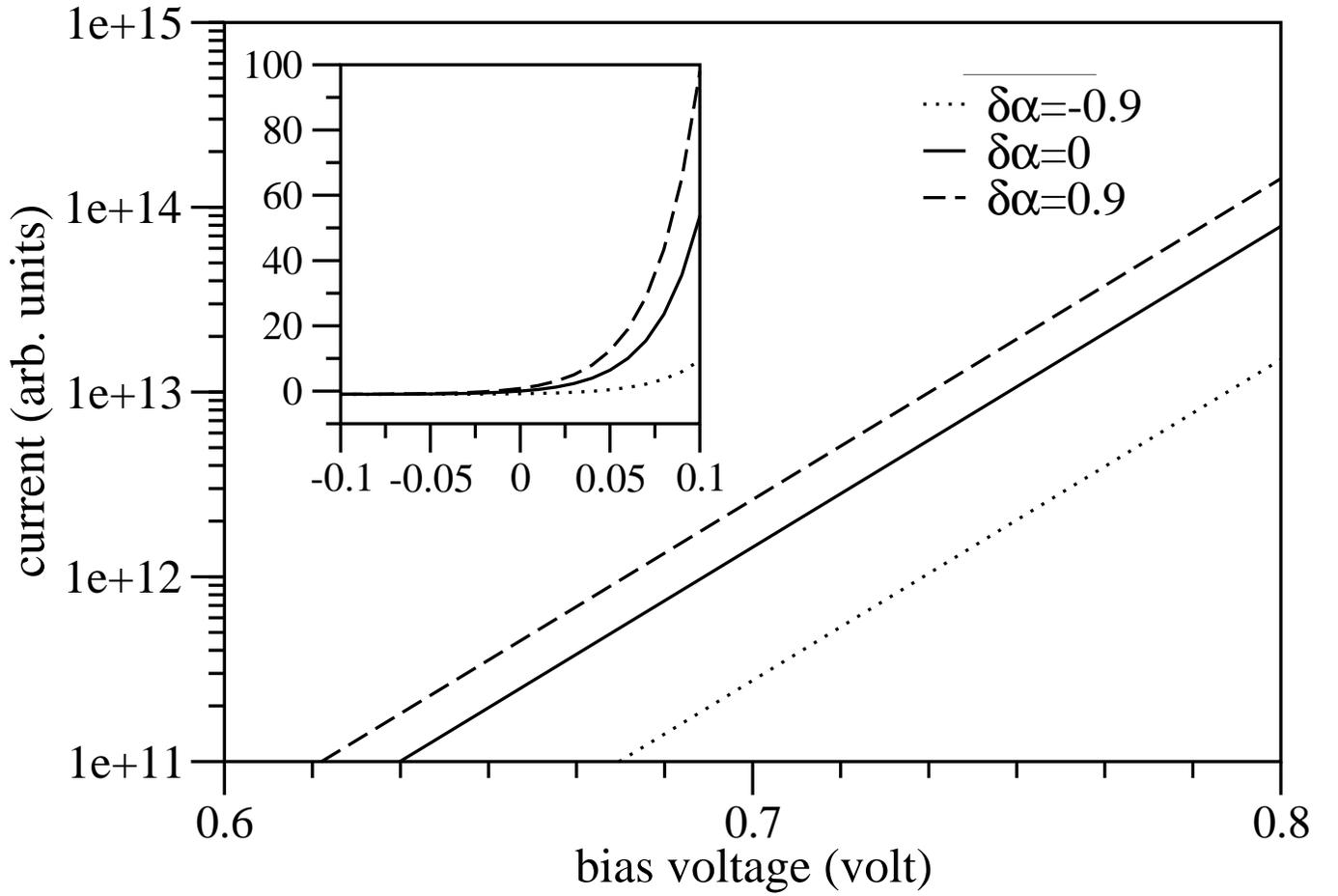}}
\caption{Calculated I-V characteristic of a magnetic {\it p-n} junction. The
current is in arbitrary units, the bias voltage in volts.  The three curves
are for different nonequilibrium spin polarizations $\delta \alpha$ at the
depletion layer in the nonmagnetic $n$ region. The equilibrium spin polarization
of the magnetic $p$ region is $\alpha_{0}=0.9$. For a fixed forward bias,
the current increases with increasing $\delta \alpha$, for the values shown
by one decade. If 
either $\delta \alpha$ or $\alpha_0$ are 
flipped by magnetic field, this result demonstrates the giant magnetoresistance
effect with the relative change of the current upon magnetic field reversal
of about 1000\%.  The inset shows the region of the I-V characteristic
around zero bias. At exactly $V=0$, the current flows for both $\delta \alpha=-0.9$
and $\delta \alpha=0.9$, (equal in magnitude but of opposite sign) demonstrating the
spin-voltaic and spin-valve effects in magnetic {\it p-n} junctions. 
}
\label{fig:5}
\end{figure}

More important is the effect of source spin on the I-V characteristic of the junction.
By applying forward bias, the spin-charge coupling is magnified (exponentially
in ideal junction diodes), leading to large giant magnetoresistance of the junction.
We illustrate here this effect with the help of our analytic formula derived using the
model introduced in Ref.~\cite{fabian02b}. The charge current in the junction with nonequilibrium
source spin polarization $\delta \alpha$ in the $n$ side of the depletion layer and
the equilibrium spin polarization $\alpha_0$ on the $p$ side of the junction is
\begin{equation}
I=I_0\left [e^V(1+\alpha_0 \delta \alpha)-1   \right ],
\end{equation}
where $I_0$ is the generation current of the diode (a material constant, which, however,
depends on magnetic field through the dependence of the equilibrium minority carrier
density on it). In Fig.~\ref{fig:5} we plot the I-V characteristic (with the current rescaled
by $I_0$) for a magnetic {\it p-n} junction with $\alpha_0=0.9$, and the source
spin polarizations $\delta \alpha_0=-0.9$, 0, and 0.9. Giant magnetoresistance is
manifest by changes in the magnitude of the current upon changing the relative sign
of $\delta \alpha$ and $\alpha_0$. In the example shown, the change in the current is
one order of magnitude, which is a relative change of 1000\%. Giant magnetoresistance
would further increase with the magnitudes of $\delta \alpha$ and $\alpha_0$. We
do not see fundamental reasons why such large magnetoresistance changes (to be
compared to typically 
30-50\% in metal-based giant magnetoresistance and 25\% in all-semiconductor 
structures~\cite{schmidt01}) would not be reproduced.


\section{Conclusions}
We have reviewed the fundamental concepts of spin-charge coupling and of bipolar spintronics. 
We have formulated spin-polarized  bipolar 
transport in inhomogeneously doped magnetic semiconductors
as carrier (electron and hole) recombination and spin relaxation limited drift and
diffusion, and discussed the solutions--both numerical and analytic--of these equations
for particular cases. The first important case was a nonmagnetic but spin-polarized
{\it p-n} junction, with spin introduced into the minority region. We have discussed
how the spin can be injected through the depletion layer leading to spin pumping in
the majority side and to spin amplification. The other device scheme where our
equations can be solved to predict the charge response to nonequilibrium spin 
is magnetic diode, in which magnetic doping in one of the two regions is inhomogeneous. 
The diode can serve to inject spin from the magnetic majority region only at large biases where 
nonequilibrium spin develops first in the magnetic region. The most
striking phenomena of the spin-voltaic effects, spin-valve effects, and giant 
magnetoresistance appear in magnetic {\it p-n} junctions if nonequilibrium spin
is externally injected to the majority nonmagnetic region, and the minority region is 
magnetic. The predicted phenomena (particularly giant magnetoresistance with
possibly more than 1000\% efficiency) should be technologically useful for sensing
magnetic fields and, more generally, for injecting, manipulating, and storing 
spin in spintronic devices. 

This work was supported by DARPA, the US ONR, and the NSF-ECS.



\newpage

\newpage
\newpage

\begin{figure}
\centerline{\psfig{file=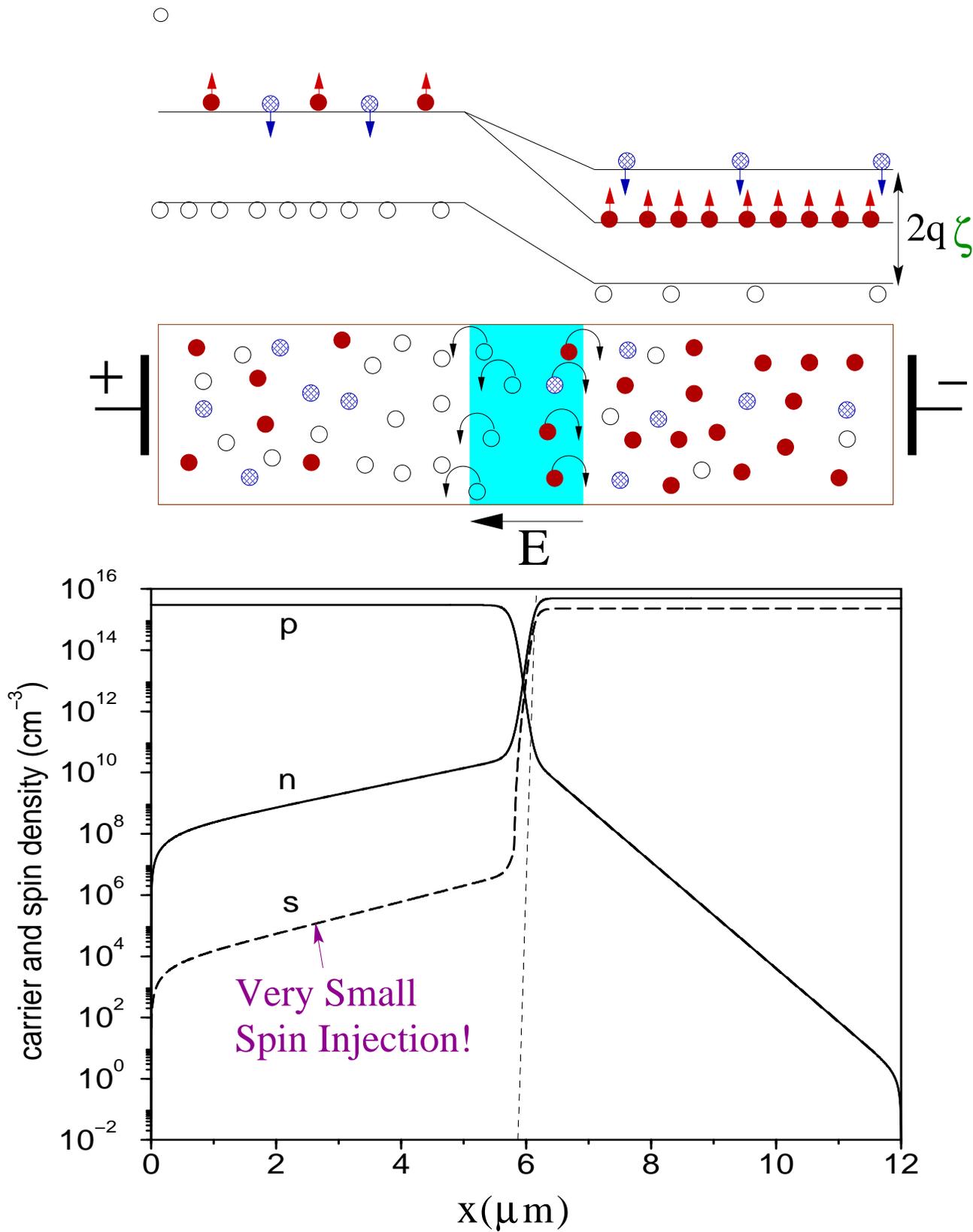,width=1\linewidth,angle=0}}
\caption{A magnetic {\it p-n} junction is a {\it p-n} junction in which
one or both regions are magnetic, with an equilibrium electron
(the case here) or hole spin polarization. A forward bias drives electrons
from the magnetic majority region to the nonmagnetic minority region, 
but no spin injection is possible at low biases, as demonstrated by 
a realistic calculation on a GaAs-based diode shown below. The graph shows the
electron ($n$), hole ($p$), and spin ($s$) densities for a 12 $\mu$m
long GaAs diode with the equilibrium spin on the $n$ side. Only a residual
spin appears as a result of a reasonably strong forward bias. The reason
for the absence of spin injection through the depletion layer is the
fact that while there are indeed more, say,  spin-up than spin-down
electrons in  the magnetic region, the barrier heights for crossing to 
the minority side are higher for the more populated spin-up species,
exactly balancing their greater population density.
Schematic representation of electron and holes
follows the notation from Fig.~\ref{fig:1}. 
}
\label{fig:3}
\end{figure}

\newpage

\begin{figure}
\centerline{\psfig{file=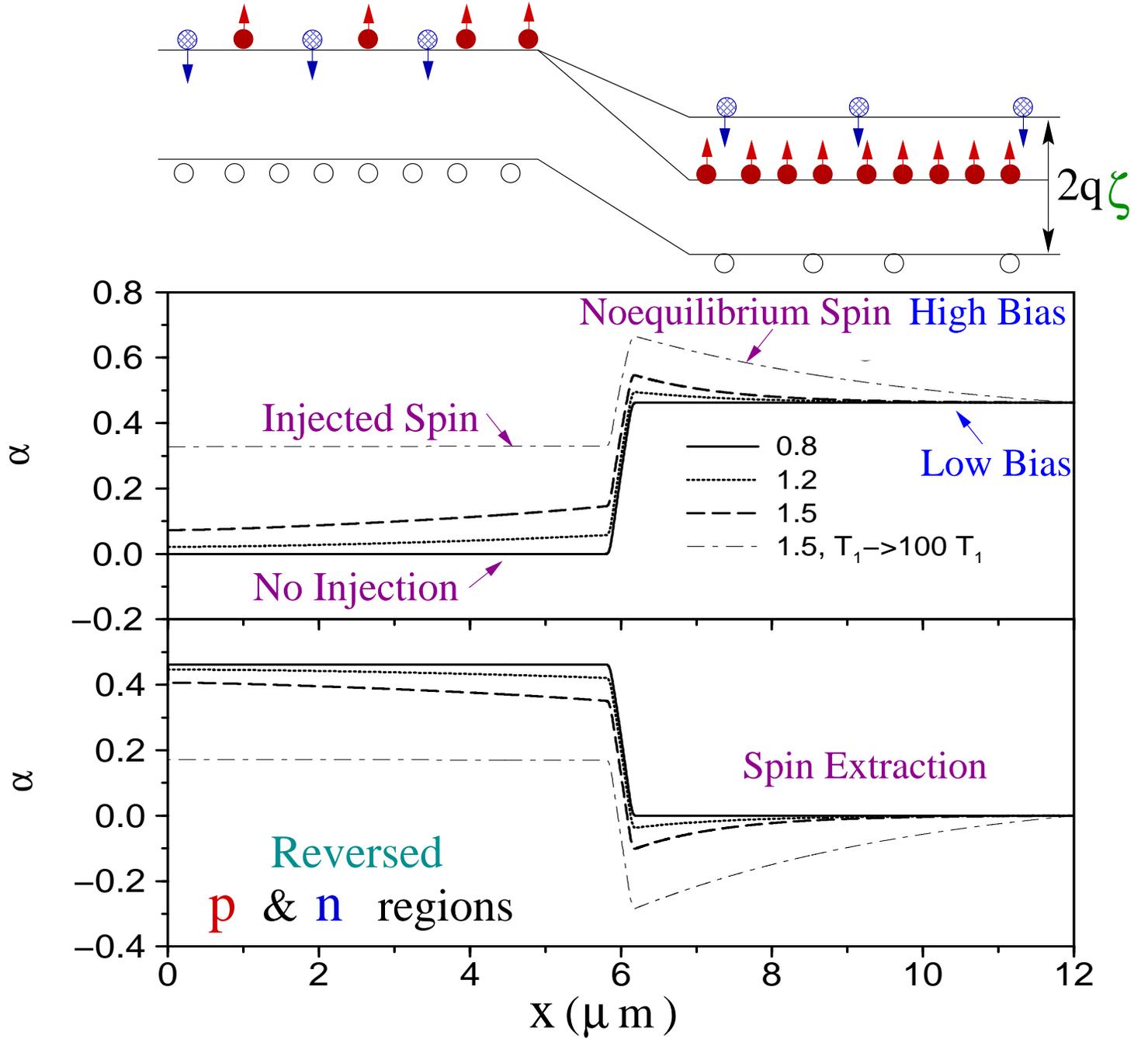,width=1\linewidth,angle=-90}}
\caption{Spin injection through magnetic {\it p-n} junctions is 
possible  
once the nonequilibrium spin appears at the junction. This
can be achieved by an external spin injection into the magnetic region
(here the $n$ region as in Fig.~\ref{fig:3}), or by applying 
large biases. The latter situation is demonstrated in the upper
graph using a calculation of the diode as in Fig.~\ref{fig:3}. 
Shown are the spin polarization $\alpha$ profiles for different forward
biases. No spin injection is seen at 0.8 volts, but spin is injected
at biases of 1.2 volt and greater. Spin injection becomes more effective
at these large biases if spin relaxation time $T_1$ increases. Together
with the appearance of the injected spin in the $p$ side, a nonequilibrium
spin appears at the depletion layer in the magnetic side. Analogous but
opposite situation occurs for reverse bias (the lower graph), where spin
becomes extracted from the nonmagnetic region $n$ region, and consequently depleted
from the magnetic $p$ region. 
Schematic representation of electron and holes
follows the notation from Fig.~\ref{fig:1}. 
}
\label{fig:4}
\end{figure}

\newpage

\begin{thebibliography}{}

\bibitem{dassarma01} 
S. Das Sarma, J. Fabian, X. Hu, and I. \v{Z}uti\'{c},
Solid State Commun. {\bf 119}, 207 (2001).

\bibitem{dassarma00a} 
S. Das Sarma, J. Fabian, X. Hu, and I. \v{Z}uti\'{c},
IEEE Transaction on Magnetics {\bf 36}, 2821 (2000).

\bibitem{dassarma00b} 
S. Das Sarma, J. Fabian, X. Hu, and I. \v{Z}uti\'{c},
Superlattice Microst.  {\bf 27}, 289 (2000).

\bibitem{optical84}
F. Meier and B. P. Zakharchenya (eds),
{\it Optical Orientation} (North-Holland, New York 1984).          

\bibitem{fabian99}
J. Fabian and S. Das Sarma, 
J. Vac. Sci. Technol. B {\bf 17}, 1780 (1999).                                                            
\bibitem{CdMnGeP}
G. A. Medvedkin, T. Ishibashi, T. Nishi, K. Hayata, Y. Hasegawa,
and K. Sato, Jpn. J. Appl. Phys. {\bf 39}, L949 (2000).                      

\bibitem{reed02}
M. L. Reed, N. A. El-Masry, H. H. Stadelmaier, Appl. Phys. Lett.
79, 3473 (2001).
                                                                                        
\bibitem{hebard02}
N. Theodoropoulou, A. F. Hebard, M. E. Overberg, C. R. Abernathy,
S. J. Pearton, S. N. G. Chu, and R. G. Wilson,
cond-mat/0201492.                                                              

\bibitem{matsumoto01}
Y. Matsumoto, M. Murakami, T. Shono, T. Hasegawa, T. Fukumura,
M. Kawasaki, P. Ahmet, T. Chikyow, S. Koshihara, and  H. Koinuma,
{\sl Science} {\bf 291}, 854 (2001).                                                    

\bibitem{cho02} 
S. Cho, S. Choi, G.-B. Cha, S. C. Hong, Y. Kim, Y.-J. Zhao, A. J. Freeman, 
J. B. Ketterson, B. J. Kim, Y. C. Kim, and B.-C. Choi,
Phys. Rev. Lett. {\bf 88}, 257203 (2002).

\bibitem{koshihara97}
S. Koshihara, A. Oiwa, M. Hirasawa, S. Katsumoto,
Y. Iye, S. Urano, H. Takagi, and H. Munekata,
Phys. Rev. Lett. {\bf 78}, 4617 (1997).

\bibitem{oiwa02}
A. Oiwa, Y. Mitsumori, R. Moriya, T. Supinski, and H. Munekata,
Phys. Rev. Lett. {\bf 88}, 137202 (2002).  
 
\bibitem{ohno00}
H. Ohno, D. Chiba, F. Matsukura, T. Omiya,
E. Abe, T. Dietl, Y. Ohno, and K. Ohtani, {\sl Nature}
{\bf 408}, 944 (2000).                                             

\bibitem{park02}
Y. D. Park, A. T. Hanbicki, S. C. Erwin, C. S. Hellberg,
J. M. Sullivan, J. E. Mattson, T. F. Ambrose, A. Wilson, G. Spanos, 
and B. T. Jonker, {\sl Science} {\bf 295}, 651 (2002).

\bibitem{rashba02a}
E. I. Rashba, J. Supercond. {\bf 15}, 13 (2002).

\bibitem{zutic02}
I. \v{Z}uti\'{c}, J. Supercond. {\bf 15}, 5 (2002).

\bibitem{rashba60}
E. I. Rashba,
Sov. Phys. Solid State {\bf 2}, 1109 (1960).
 
\bibitem{rashba84}
Y. A. Bychkov and E. I. Rashba, J. Phys. C {\bf 17}, 6039 (1984).
 
\bibitem{nitta97}
J. Nitta, T. Akazaki, H. Takayanagi, and T. Enoki, 
Phys. Rev. Lett. {\bf 78}, 1335 (1997).

\bibitem{miller02}
J. B. Miller, D. H. Zumb\"{u}hl, C. M. Marcus, Y. B. Lyanda-Geller,
D. Goldhaber-Gordon, K. Campman, and A. C. Gossard, 
cond-mat/0206375.

\bibitem{gorkov01}
L. P. Gor'kov and E. I. Rashba,
Phys. Rev. Lett. {\bf 87}, 037004 (2001).

\bibitem{hausler01}
W. H\"{a}usler, Phys. Rev. B {\bf 63}, 121310 (2001).

\bibitem{aronov76a}
A. G. Aronov and G. E. Pikus,
Sov. Phys. Semicond. {\bf 10}, 698 (1976).                

\bibitem{aronov76b}
A. G. Aronov, JETP Lett. {\bf 24}, 32 (1976).                                 

\bibitem{silsbee80}
R. H. Silsbee, Bull. Magn. Reson. {\bf 2}, 284 (1980).
 
\bibitem{johnson85}
M. Johnson and R. H. Silsbee, 
Phys.  Rev. Lett. {\bf 55}, 1790 (1985).

\bibitem{johnson87}
M. Johnson and R. H. Silsbee, 
Phys.  Rev. B {\bf 35}, 4959 (1987).

\bibitem{datta90}
S. Datta and B. Das, Appl. Phys. Lett. {\bf 56} 665 (1990).

\bibitem{dassarma00c}
S. Das Sarma, J. Fabian, X. Hu, and I. \v{Z}uti\'{c},
58th DRC (Device Research Conference) Conference Digest,
p. 95-8 (IEEE, Piscataway, 2000); cond-mat/0006369.       

\bibitem{zutic01a} 
I. \v{Z}uti\'{c}, J. Fabian, and S. Das Sarma,
Phys. Rev. B {\bf 64}, 121201 (2001).

\bibitem{zutic01b} 
I. \v{Z}uti\'{c}, J. Fabian, and S. Das Sarma,
Appl. Phys. Lett. {\bf 79}, 1558 (2001).

\bibitem{zutic02a} 
I. \v{Z}uti\'{c}, J. Fabian, and S. Das Sarma,
Phys. Rev. Lett. {\bf 88},  066603 (2002).

\bibitem{zutic02b} 
I. \v{Z}uti\'{c}, J. Fabian, and S. Das Sarma,
cond-mat/0205177.

\bibitem{fabian02a} 
J. Fabian, I. \v{Z}uti\'{c}, and S. Das Sarma,
cond-mat/0205340.

\bibitem{rashba02b}
E. I. Rashba, cond-mat/0206129.

\bibitem{landau36}
L. D. Landau and E. M. Lifshitz, 
Phys. Zs. Sowjet. {\bf 9}, 477 (1936).

\bibitem{tiwari92} 
S. Tiwari, {\it Compound Semiconductor Device Physics} 
(Academic Press, San Diego, 1992).

\bibitem{martin02}
I. Martin, preprint cond-mat/0201481.                              

\bibitem{yu02}
Z. G. Yu and M. E. Flatt\'{e}, preprint cond-mat/0201425.

\bibitem{fabian02b} 
J. Fabian, I. \v{Z}uti\'{c}, and S. Das Sarma,
unpublished. 

\bibitem{furdyna88}
J. K. Furdyna and J. Kossut,
{\it Semiconductors and Semimetals}, Vol. 25,
(Academic, New York, 1988).                                                  

\bibitem{dietl94}  
T. Dietl, in {\it Handbook of Semiconductors} Vol. 3, 
edited by T. S. Moss and S. Mahajan, p. 1279 
(Noth-Holland, New York, 1994). 

\bibitem{fiederling99} 
R. Fiederling, M. Kleim, G. Reuscher, W. Ossau, 
G. Schmidt, A. Waag, and L. W.  Molenkamp, 
{\sl Nature} {\bf 402}, 787 (1999).

\bibitem{jonker00} 
B. T. Jonker, Y. D. Park, B. R. Bennett, H. D. Cheong,
G. Kioseoglou, and A. Petrou, Phys. Rev. B {\bf 62}, 
8180 (2000).

 
\bibitem{schmidt01}
G. Schmidt, G. Richter, P. Grabs, C. Gould, D. Ferrand, and L. W. Molenkamp,
Phys. Rev. Lett. {\bf 87}, 227203 (2001).


\end{thebibliography}
\end{document}